\documentclass{llncs}
\pdfoutput=1
\usepackage{amsxtra}
\usepackage{times}
\usepackage{mathpartir}
\usepackage{mathtools}
\usepackage{tikz}
\usepackage{enumerate}
\usepackage{diagrams}

\newcounter{transcount}

\newcommand\mytik{\begingroup \catcode`\&13 \tikmy}
\newcommand{\tikmy}[1]{
 \setcounter{transcount}{0}
\begin{tikzpicture} [
  column 1/.style={anchor=west},
  column 2/.style={anchor=west},
  column 3/.style={anchor=west},
  column sep= 0cm]
  {#1}
\end{tikzpicture}
\endgroup}

\newcommand{\tc}[1]{
 \addtocounter{transcount}{1}
 \ifodd #1
 (\arabic{transcount})
 \else
 (\roman{transcount})
 \fi}

\newcommand{\rstc}{\setcounter{transcount}{0}}

\newcommand{\lts}[1]{\mathbin{\xrightarrow{#1}}}
\newcommand{\Lts}[1]{\mathbin{\xRightarrow{#1}}}

\newcommand{\E}{{\mathcal{E}}}
\newcommand{\snd}[1]{{#1}.2}
\newcommand{\fst}[1]{{#1}.1}
\newcommand{\X}{{\mathcal{X}}}
\newcommand{\Y}{{\mathcal{Y}}}

\newcommand{\geb}[6]{{#3\mathbin{#2^{#1}_{#6;#5}}#4}}
\newcommand{\ceb}[0]{\geb{\star}}
\newcommand{\meb}[0]{\geb{-}}
\newcommand{\eb}[0]{\geb{}}
\newcommand{\EB}[1]{\eb{\X}{#1}}
\newcommand{\Eq}[1]{\eb{\sim}{#1}}
\newcommand{\cEB}[1]{\ceb{\X}{#1}}
\newcommand{\cEBY}[1]{\ceb{\Y}{#1}}
\newcommand{\mEBY}[1]{\meb{\Y}{#1}}

\newcommand{\gClos}[2]{\mathbin{(#2;#1)^\star}}
\newcommand{\Clos}{\gClos{r}{\E}}
\newcommand{\rClos}[1]{\gClos{#1}{\E}}
\newcommand{\eClos}[1]{\gClos{r}{#1}}
\newcommand{\zClos}{\eClos{\emptyset}}

\newcommand{\gpClos}[2]{\mathbin{(#2;#1)^\circ}}
\newcommand{\pClos}{\gpClos{r}{\E}}
\newcommand{\rpClos}[1]{(\E;#1)^\circ}
\newcommand{\epClos}[1]{(#1;r)^\circ}

\newcommand{\ba}[1]{{\overline{#1}}}
\newcommand{\ti}[1]{{\widetilde{#1}}}
\newcommand{\M}{\ti M}
\newcommand{\N}{\ti N}

\newcommand{\adj}[2]{#1{\oplus}#2}

\newcommand{\prl}{\mathbin{|}}
\newcommand{\bang}{{!}}

\newcommand{\eqv}{\equiv}

\definecolor{darkgreen}{RGB}{0,128,0}

\newcommand{\lock}{\mathit{lock}}
\newcommand{\unlock}{\mathit{unlock}}

\newcommand{\simcong}{\mathbin{\approx_{c}}}

\newcommand{\mytitle}{Sound Bisimulations for Higher-Order Distributed Process Calculus}

\newcommand{\resp}{resp.\ }
\newcommand{\run}{\textit{run}}

\newcommand{\proofsketch}[1]{\emph{Proof sketch.}}

\begin{document}

% LATEXMAKE gappei.pdf
% LATEXEXTRA
%
\title{\mytitle\thanks{Appendix with full proofs at \url{http://www.kb.ecei.tohoku.ac.jp/\~adrien/pubs/SoundAppendix.pdf}}}

\author{Adrien Pi\'erard and Eijiro Sumii\thanks{This research is partially supported by
KAKENHI 22300005, the Nakajima Foundation, and the Casio Science Promotion Foundation. The first author is
partially supported by the Global COE Program CERIES.}}
\institute{
Tohoku University\\
\email{\{adrien,sumii\}@kb.ecei.tohoku.ac.jp}
}

\maketitle

% LATEXMAKE gappei.tex
% LATEXEXTRA -- vim extra
% vim: set textwidth=90 spelllang=en_gb spell:
\begin{abstract}
  While distributed systems with transfer of processes have become pervasive, methods for
  reasoning about their behaviour are underdeveloped. In this paper we propose a
  bisimulation technique for proving behavioural equivalence of such systems modelled in
  the \emph{higher-order $\pi$-calculus with passivation} (and restriction).  Previous
  research for this calculus is limited to context bisimulations and normal bisimulations
  which are either impractical or unsound. In contrast, we provide a sound and useful
  definition of \emph{environmental bisimulations}, with several non-trivial
  examples.  Technically, a central point in our bisimulations is the clause for parallel
  composition, which must account for passivation of the spawned processes in the middle
  of their execution.
\end{abstract}

% LATEXMAKE gappei.tex
% LATEXEXTRA -- vim extra
% vim: set textwidth=90 spelllang=en_gb spell:
\section{Introduction}
\label{sec:intro}
\subsection{Background}
Higher-order distributed systems are ubiquitous in today's computing environment. To name
but a few examples, companies like Dell and Hewlett-Packard sell products using virtual
machine live migration~\cite{dell,hp}, and Gmail users execute remote JavaScript
code on local browsers.  In this paper we call \emph{higher-order} the ability to transfer
processes, and \emph{distribution} the possibility of location-dependent system behaviour.
In spite of the \emph{de facto} importance of such systems, they are hard to analyse
because of their inherent complexity.

The $\pi$-calculus~\cite{pi} and its dialects prevail as models of concurrency, and
several variations of these calculi have been designed for distribution. First-order
variations include the ambient calculus~\cite{ambients} and D$\pi$~\cite{dpi}, while
higher-order include more recent Homer~\cite{homer} and Kell~\cite{kell} calculi. In this
paper, we focus on the higher-order $\pi$-calculus with \emph{passivation}~\cite{hopip}, a
simple high-level construct to express distribution.  It is an extension of the
higher-order $\pi$-calculus~\cite{hop} (with which the reader is assumed to be familiar)
with \emph{located processes} $a[P]$ and two additional transition rules:
$a[P] \lts{\ba a\langle P\rangle} 0$ (\textsc{Passiv}), and
$a[P] \lts{\alpha} a[P']$ if $P \lts{\alpha} P'$ (\textsc{Transp}).

The new syntax $a[P]$ reads as ``process $P$ located at $a$'' where $a$ is a name.  Rule
\textsc{Transp} specifies the transparency of locations, i.e.\ that a location has no
impact on the transitions of the located process.  Rule \textsc{Passiv} indicates that a
located process can be \emph{passivated}, that is, be output to a channel of the same name
as the location.  Using passivation, various characteristics of distributed systems are
expressible.  For instance, failure of process $P$ located at $a$ can be modelled like
$a[P] \prl a(X).\ba{\mathit{fail}} \lts{} 0\prl\ba{\mathit{fail}}$, and migration of
process $Q$ from location $b$ to $c$ like $b[P] \prl b(X).c[X] \lts{} 0 \prl c[P]$.

One way to analyse the behaviour of systems is to compare implementations and
specifications. Such comparison calls for satisfying notions of behavioural equivalence,
such as \emph{reduction-closed barbed equivalence}
(and \emph{congruence})
~\cite{honda},
written $\approx$ (and $\simcong$ respectively) in this paper.

Unfortunately, these equivalences have succinct definitions that are not very practical as a proof technique,
for they both include a condition that quantifies over arbitrary processes, like: if
$P\approx Q \text{ then } \forall R.~ P \prl R \approx Q \prl R $.  Therefore, more
convenient definitions like \emph{bisimulations}, for which membership implies behavioural
equivalence, and which come with a co-inductive proof method, are sought after.

Still, the combination of both higher order and distribution has long been considered
difficult. Recent research on higher-order process calculi led to defining sound
\emph{context bisimulations}~\cite{context} (often at the cost of appealing to Howe's
method~\cite{Howe} for proving congruence) but those bisimulations suffer from their heavy
use of universal quantification: suppose that $\nu \ti c.\ba a \langle M \rangle .P
\mathbin{\X}   \nu \ti d.\ba a \langle N\rangle .Q,$ where $\X$ is a context bisimulation;
then it is roughly required that
for any process $R$,
 we have $\nu \ti c.(P\prl R\{M/X\}) \mathbin{\X} \nu \ti d.(Q\prl R\{N/X\})$.
Not only must we consider the outputs $M$ and $N$, but we must also handle interactions of
arbitrary $R$ with the continuation processes $P$ and $Q$. Alas, this almost comes down to
showing reduction-closed barbed equivalence!  In the higher-order $\pi$-calculus, by means
of encoding into a first-order calculus, normal bisimulations~\cite{context} coincide with
(and are a practical alternative to) context bisimulations.  Unfortunately, normal
bisimulations have proved to be unsound in the presence of passivation (and
restriction)~\cite{hopip}.
While this result cast a doubt on whether sound normal
bisimulations exist for higher-order distributed calculi, it did not affect the potential
of environmental bisimulations ~\cite{sealing,typeabs,EB,hoapp} as a useful proof
technique for behavioural equivalence  in those calculi.

\subsection{Our contribution}
To the best of our knowledge, there are not yet any useful sound
bisimulations for higher-order distributed process calculi. In this paper we develop
environmental (weak) bisimulations for the higher-order $\pi$-calculus with passivation,
which (1) are sound with respect to reduction-closed barbed equivalence, (2)
can actually be used to prove behavioural equivalence of non-trivial processes (with
restrictions), and (3) can also be used to prove reduction-closed barbed \emph{congruence} of
processes (see Corollary~\ref{cor:congruence}).  To prove reduction-closed barbed
equivalence (and congruence), we find a new clause to guarantee preservation of
bisimilarity by parallel composition of arbitrary processes. Unlike the corresponding
clause in previous research~\cite{hopip,hoapp}, it can also handle the later removal
(i.e.\ passivation) of these processes while keeping the bisimulation proofs tractable.
Several examples are given, thereby supporting our claim of the first useful bisimulations
for a higher-order distributed process calculus.  Moreover, we define an up-to context
variant of the environmental bisimulations that significantly lightens the burden of
equivalence proofs, as utilised in the examples.

\paragraph*{Overview of the bisimulation} %
\label{ssub:verview of the bisimulation}
We now outline the definition of our environmental bisimulations.  (Generalities
on environmental bisimulations can be found in~\cite{EB}.) We define an
environmental bisimulation $\X$ as a set of quadruples $(r,\E,P,Q)$ where $r$ is a set of
names (i.e.\ channels and locations), $\E$ is a binary relation (called the
\emph{environment}) on terms, and $P$, $Q$ are processes. The bisimulation is a game where
the processes $P$ and $Q$ are compared to each other by an \emph{attacker} (or
\emph{observer}) who knows and can use the terms in the environment $\E$ and the names in
$r$. For readability, the membership $(r,\E,P,Q)\in \X$ is often written
$\EB{P}{Q}{r}{\E}$, and should be understood as ``processes $P$ and $Q$ are bisimilar,
under the environment $\E$ and the known names $r$.''

The environmental bisimilarity is co-inductively defined by several conditions concerning
the tested processes and the knowledge. As usual with weak bisimulations, we require that
an internal transition by one of the processes is matched by zero or more internal
transitions by the other, and that the remnants are still bisimilar.

As usual with (more recent and less common) environmental bisimulations, we require that
whenever a term $M$ is output to a known channel, the other tested process can output
another term $N$ to the same channel, and that the residues are bisimilar under the
environment extended with the pair $(M,N)$. The extension of the environment stands for
the growth of knowledge of the attacker of the bisimulation game who observed the outputs
$(M,N)$, although he cannot analyse them.
This spells out like:
for any $\EB{P}{Q}{r}{\E}$ and $a\in r$,
if $P \lts{\nu \ti c.\ba a\langle M\rangle }P'$  for fresh $\ti c$, then
$Q\Lts{\nu \ti d.\ba a\langle N\rangle}Q'$ for fresh $\ti d$ and
$\EB{P'}{Q'}{r}{\E\cup\{(M,N)\}}$.

Unsurprisingly, input must be doable on the same known channel by each process, and the
continuations must still be bisimilar under the same environment since nothing is learnt
by the context.  However, we require that the input terms are generated from the
\emph{context closure} of the environment.  Intuitively, this closure represents all the
processes an attacker can build by combining what he has learnt from previous outputs.
Roughly, we define it as:
\vskip3pt
\centerline{
\(
\Clos = \{(C[\M],C[\N])\mid C\mathit{\ context},~\mathit{fn}(C)\subseteq r,~\M\mathbin{\E}\N\}
\)
}
\vskip3pt
\noindent
where $\M$ denotes a sequence $M_0,\dots,M_n$, and $~\M\E\N$ means that for all $0\leq
i\leq n$, $M_i\mathbin{\E} N_i$. Therefore, the input clause looks like:
\(
\text{for any } \EB{P}{Q}{r}{\E},~a\in r \text{ and } (M,N)\in \Clos,
\text{ if }P \lts{a(M)}P',\text{ then } Q\Lts{a(N)}Q'\text{ and }
\EB{P'}{Q'}{r}{\E}.
\)

The set $r$ of known names can be extended at will by the observer, provided that the new
names are fresh:
\(
\text{for any } \EB{P}{Q}{r}{\E} \text{ and } n \text { fresh, we have }
\EB{P}{Q}{r\cup\{n\}}{\E}.
\)

\paragraph*{Parallel composition}

The last clause is crucial to the soundness and usefulness of environmental bisimulations
for languages with passivation, and not as straightforward as the other clauses. The idea
at its base is that not only may an observer run arbitrary processes $R$ in parallel to
the tested ones (as in reduction-closed barbed equivalence), but he may also run
arbitrary processes $M,N$ he assembled from previous observations. It is critical to
ensure that bisimilarity (and hopefully equivalence) is preserved by such parallel
composition, and that this property can be easily proved.  As $\Clos$ is this set of
processes that can be assembled from previous observations, we would naively expect the
appropriate clause to look like:
\vskip3pt
\centerline{
\(
\text{For any } \EB{P}{Q}{r}{\E}\text{ and } (M,N)\in \Clos,
\text{ we have } \EB{P \prl M}{Q \prl N}{r}{\E}
\)
}
\vskip3pt
\noindent
but this subsumes the already impractical clause of reduction-closed barbed equivalence
which we want to get round.  Previous research~\cite{hopip,hoapp} uses a weaker condition:
\vskip3pt
\centerline{
\(
\text{For any } \EB{P}{Q}{r}{\E}\text{ and } (M,N)\in \E,
\text{ we have } \EB{P \prl M}{Q \prl N}{r}{\E}
\)
}
\vskip3pt
\noindent
arguing that $\Clos$ can informally do no more observations than $\E$, but this clause is
unsound in the presence of passivation. The reason behind the unsoundness is that, in
our settings, not only can a context spawn new processes $M$, $N$, but it can also
\emph{remove} running processes it created by passivating them later on.  For example, consider
the following processes $P = \ba a \langle R \rangle . ! R$ and $Q = \ba a \langle 0
\rangle . ! R$. Under the above weak condition, it would be easy to construct an
environmental bisimulation that relates $P$ and $Q$. However, a process $a(X).m[X]$ may
distinguish them. Indeed, it may receive processes $R$ and start running it in location
$m$, or may receive process $0$ and run a copy of $R$ from $\bang R$. If $R$ is a process
doing several sequential actions (for example if $R = \lock.\unlock$) and is passivated
\emph{in the middle} of its execution, then the remaining processes after passivation
would not be equivalent any more.

To account for this new situation, we decide to modify the condition on the provenance of
process that can be spawned, drawing them from $\{(a[M],a[N]) \mid a\in r,\ (M,N)\in\E\}$,
thus giving the clause:
\vskip3pt
\centerline{
\(
\text{For any } \EB{P}{Q}{r}{\E},~a\in r \text{ and } (M,N)\in \E,
\text{ we have } \EB{P \prl a[M]}{Q \prl a[N]}{r}{\E}.
\)
}
\vskip3pt
\noindent
The new condition allows for any running process that has been previously created by the
observer to be passivated, that is, removed from the current test.
This clause is much more tractable than the first one using $\Clos$ and, unlike
the second one using only $\E$, leads to sound environmental bisimulations
(albeit with a limitation; see Remark~\ref{rem:simple}).

\paragraph*{Example}
\label{ssub:Example}
With our environmental bisimulations, non-trivial equivalence of higher-order distributed
processes can be shown, such as $P_0 =\bang{}a[e \prl \ba e]$ and $Q_0 = \bang{}a[e] \prl
\bang{}a[\ba e]$, where $e$ abbreviates $e(X).0$ and $\ba e$ is $\ba e\langle 0\rangle.0$.
We explain here informally how we build a bisimulation $\X$ relating those processes.
\[
\begin{array}{ll}
\X = \{(r,\E, P ,Q )\mid
&r\supseteq\{a,e\}, \hskip0.5em \E = \{0,e,\ba e,e \prl \ba e\} \times \{0,e,\ba e\},\\
&P \eqv P_0 \prl \prod_{i=1}^nl_i[M_i], \hskip0.5em Q \eqv Q_0 \prl \prod_{i=1}^nl_i[N_i], \hskip0.5em n\ge0,\\
&\ti l\in r,  \hskip0.5em (\M,\N)\in \E\}
\end{array}
\]
Since we want $\EB{P_0}{Q_0}{r}{\E}$, the spawning clause of the bisimulation requires
that for any $(M_1,N_1)\in \E$ and $l_1\in r$, we have $\EB{P_0 \prl l_1[M_1]}{Q_0 \prl
l_1[N_1]}{r}{\E}$. Then, by repeatedly applying this clause, we obtain $\EB{(P_0 \prl
\prod_{i=1}^nl_i[M_i])}{(Q_0 \prl \prod_{i=1}^nl_i[N_i])}{r}{\E}$.  Since the observer can
add fresh names at will, we require $r$ to be a superset of the free names $\{a,e\}$ of
$P_0$ and $Q_0$.  Also, we have the intuition that the only possible outputs from $P$ and
$Q$ are processes $e \prl \ba e$, $e$, $\ba e$, and $0$.  Thus, we set ahead $\E$ as the
Cartesian product of $\{0, e, \ba e, e \prl \ba e\}$ with $\{0, e, \ba e\}$, that is, the
combination of expectable outputs.  We emphasize that it is indeed reasonable to relate
$\ba e,e$ and $e \prl \ba e$ to $0,e$ and $\ba e$ in $\E$ for the observer cannot analyse
the pairs: he can only use them along the tested processes $P$ and $Q$ which, by the
design of environmental bisimulations, will make up for the differences.

Let us now observe the possible transitions from $P$ and their corresponding transitions
from $Q$ by glossing over two pairs of trees, where related branches represent the
correspondences.  (Simulation in the other direction is similar and omitted for brevity.)
First, let us consider the input and output actions as shown in
Figure~\ref{fig:observable}.  (i) When $P_0$ does an input action $e$ or an output action
$\ba e$, it leaves behind a process $a[\ba e]$ or $a[e]$, respectively.  $Q_0$ can also do
the same action, leaving $a[0]$.  Since both $(\ba e,0)$ and $(e,0)$ are in $\E$, we can
add the leftover processes to the respective products $\prod$; (ii) output by passivation
is trivial to match (without loss of generality, we only show the case $i=n$), and (iii)
observable actions $\alpha$ of an $M_n$, leaving a residue $M'_n$, are matched by one of
$Q_0$'s $a[\alpha]$, leaving $a[0]$.  To pair with this $a[0]$, we replicate
an $a[e \prl
\ba e]$ from $P_0$, and then, as in (i), they add up to the products $\prod$.
\newcommand{\p}{P_0}
\newcommand{\q}{Q_0}
\newcommand{\pim}[2]{\ensuremath{\prod_{i=1}^{#1}l_i[#2_i]}}
\newcommand{\pimp}[2]{\ensuremath{\prod_{i=1}^{#1}l'_i[#2_i]}}
\newcommand{\pip}[1]{\pim{#1}{M}}
\newcommand{\qip}[1]{\pim{#1}{N}}
\begin{figure}[t]
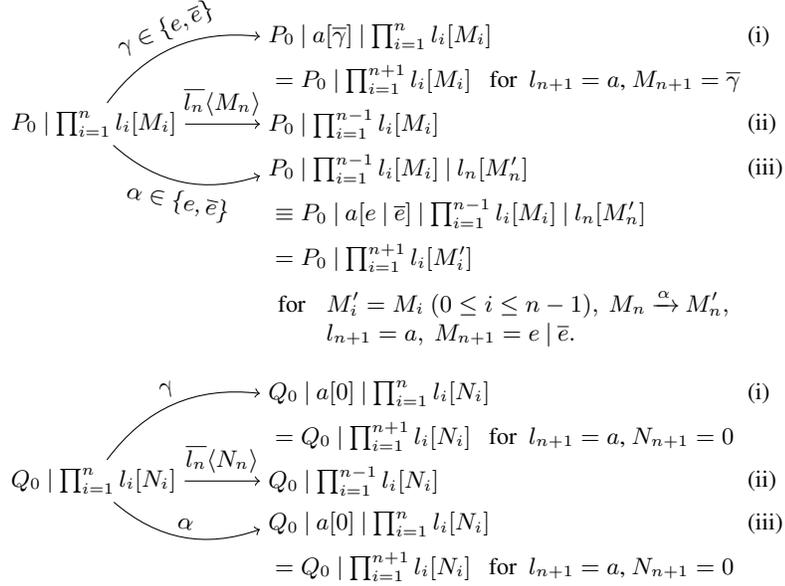

  \vspace{-0.8cm}
  
  \setlength{\belowcaptionskip}{-0.5cm}
\mytik{
  \matrix {
                                    &[1.0cm] \node(c1){$\p\prl a[\ba \gamma]\prl\pip{n}$};  &[-2mm]\node{\tc0};\\
                                    & \node{${}= \p\prl\pip{n+1}$ \text{\ }
                                    for\text{\ } $l_{n+1}= a$, $M_{n+1}=\ba \gamma$};  \\

  \node (rootp) {$\p\prl\pip{n}$};  & \node(c2){$\p\prl\pip{n-1}$};&\node{\tc0};\\

                                    & \node(c3){$\p\prl\pip{n-1}\prl l_n[M_n']$};&\node{\tc0};\\
                                    & \node    {${} \equiv \p\prl a[e \prl \ba e]\prl\pip{n-1}\prl l_n[M_n']$}; \\
                                    & \node    {${} =\p \prl\pim{n+1}{M'}$}; \\
                                    & \node    {
                                    $\begin{array}{llll}
                                      \text{ for }&M_i'=M_i~(0\leq i \le
                                      n-1),~M_n\lts{\alpha}M_n',\\
                                      & l_{n+1} = a,~M_{n+1}= e \prl \ba e.
                                    \end{array}$
                                    }; \\
    &&\\
    &&\node{\rstc};\\
    &&\\
                                    & \node(d1){$\q\prl a[0]\prl\qip{n}$};&       \node{\tc0}; \\
                                    & \node {${} = \q\prl\qip{n+1}$ \text{\ } for\text{\ } $l_{n+1}= a$, $N_{n+1}=0$};  \\
  \node (rootq) {$\q\prl\qip{n}$}; 
                                    & \node(d2){$\q\prl\qip{n-1}$};&\node{\tc0};\\

                                    & \node(d3){$\q\prl a[0] \prl\qip{n}$};&\node{\tc0};\\
                                    & \node    {${}= \q\prl\pim{n+1}{N}$ \text{\ } for\text{\ } 
                                    $l_{n+1}=a$,
                                    $N_{n+1} = 0$};\\
  };
  \draw  [->, bend left ] (rootp) to node [above,midway,sloped] {$\gamma \in \{e,\ba e\}$} (c1.west);
  \draw  [->            ] (rootp) to node [above,midway,sloped] {$\ba{l_n}\langle M_n\rangle$} (c2.west);
  \draw  [->, bend right] (rootp) to node [below,midway,sloped] {$\alpha\in \{e,\ba e\}$} (c3.base west);
  \draw  [->, bend left ] (rootq) to node [above,midway,sloped] {$\gamma $}  (d1.west);
  \draw  [->            ] (rootq) to node [above,midway,sloped] {$\ba{l_n}\langle N_n\rangle$} (d2.west) ;
  \draw  [->, bend right] (rootq) to node [above,midway,sloped] {$\alpha$}  (d3.base west);
  }
  \caption{Simulation of observable transitions}
  \label{fig:observable}
  
\end{figure}

\begin{figure}
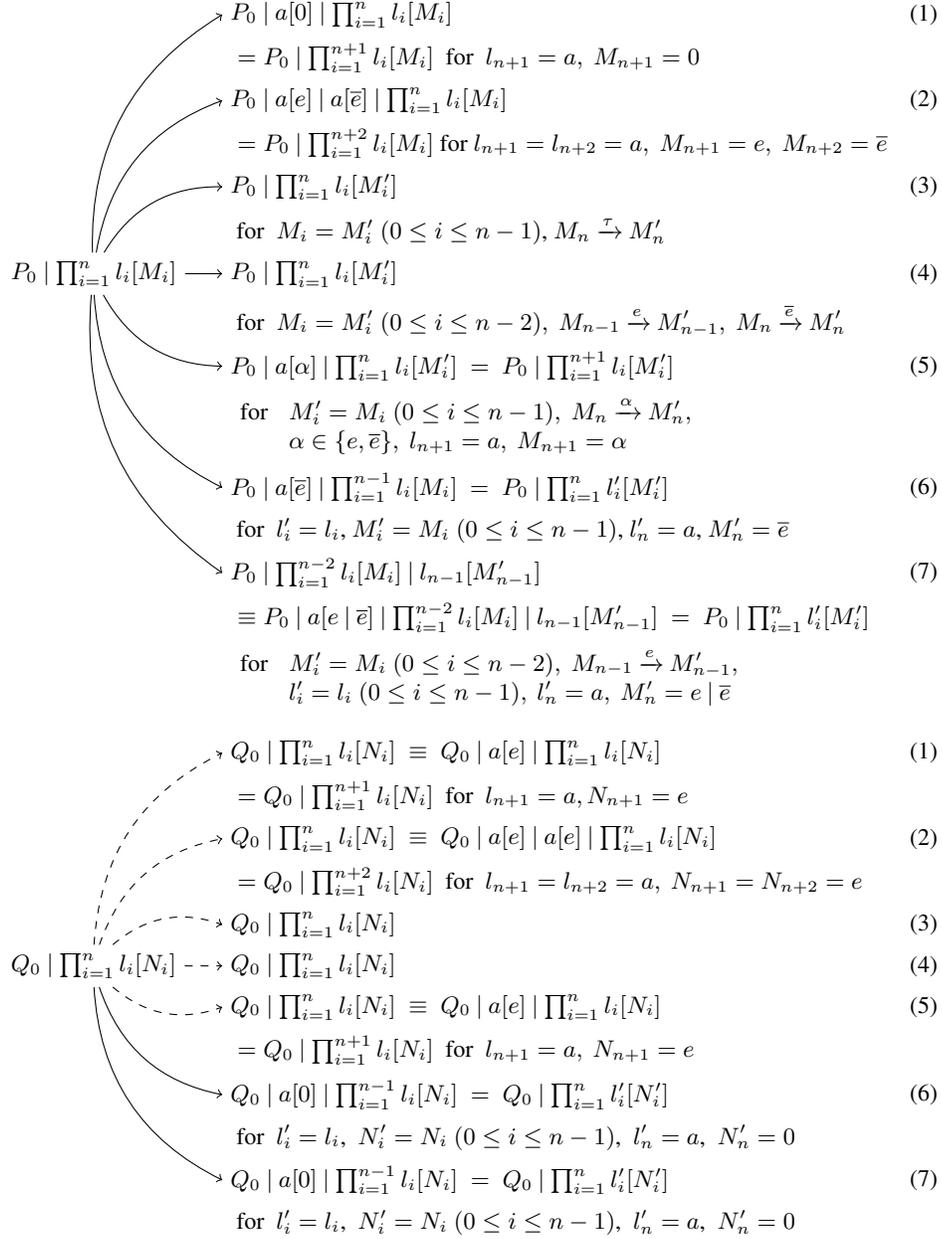

  
  \setlength{\belowcaptionskip}{-0.5cm}
\mytik{
  \matrix {

                                    &[0.5cm] \node(c1){$\p\prl a[0]\prl\pip{n}$};&\node{\tc1}; \\
                                    &\node{${}= \p\prl\pip{n+1}$ \text{ for } $l_{n+1} = a,\  M_{n+1}= 0$}; \\
                                    
                                    & \node(c2){$\p\prl a[e]\prl a[\ba e]\prl\pip{n}$};&\node{\tc1}; \\
                                    & \node    {${}=\p\prl\pip{n+2}\text{ for }
                                    l_{n+1}=l_{n+2}=a,\ 
                                    M_{n+1}=e,\ 
                                    M_{n+2}=\ba e$
                                    };\\
                                    
                                    & \node(c3){$\p\prl\pim{n}{M'}$};&\node{\tc1}; \\
                                    & \node    {\text{~for~} $M_i = M_i'~(0\le i\le n-1)$, $M_n \lts\tau M_n'$}; \\
                                    
  \node (rootp) {$\p\prl\pip{n}$};  & \node(c5){$\p\prl\pim{n}{M'}$};&\node{\tc1}; \\
                                    & \node    {\text{~for~}
                                    $M_i = M_i'~(0\le i\le n-2),\ 
                                    M_{n-1} \lts{e} M_{n-1}',\ 
                                    M_n \lts{\ba e} M_n'$
                                    };\\
                                    
                                    & \node(c7){$\p\prl a[\alpha]\prl\pim{n}{M'}\text{~}=\text{~} \p\prl\pim{n+1}{M'}$};&\node{\tc1};\\
                                    & \node    {$
                                    \begin{array}{llll}
                                      \text{ for }&M_i'=M_i~(0\le i\le n-1),\ M_n\lts{\alpha} M_n',\\
                                      &\alpha \in \{e,\ba e\},\ l_{n+1}=a,\ M_{n+1}=\alpha\\
                                    \end{array}$};\\
                                    
                                    & \node(c11){$\p\prl a[\ba e]\prl\pip{n-1} \text{~}=\text{~} \p\prl\pimp{n}{M'}$};&\node{\tc1}; \\
                                    & \node    {\text{~for~}
                                    $l_i' = l_i$,
                                    $M_i' = M_i~(0\le i\le n-1) $,
                                    $l_n' = a$,
                                    $M_n' = \ba e$
                                    };\\
                                    
                                    & \node(c12){$\p\prl \pip{n-2}\prl l_{n-1}[M_{n-1}']$};&\node{\tc1}; \\
                                    & \node     {${} \equiv \p \prl a[e \prl \ba e]\prl \pip{n-2}\prl l_{n-1}[M_{n-1}'] \text{~}=\text{~}      \p \prl \pimp{n}{M'}$}; \\
                                    & \node    {$
                                    \begin{array}{llll}
                                      \text{ for }& M_i'=M_i~(0\le i\le n-2),~M_{n-1}\lts{e}M_{n-1}',\\
                                      &l_i' = l_i~(0\le i\le n-1),~l_n' = a,~M_n' = e \prl \ba e\\
                                    \end{array}$
                                    };\\

                                    &&&&\\
                                    &&&\node{\rstc};&\\
                                    &&&&\\

                                    & \node(d1){$\q\prl \qip{n}\text{~}\equiv\text{~} \q\prl a[e]\prl\qip{n}$}; &\node{\tc1}; \\
                                    & \node {${}=\q\prl\pim{n+1}{N}\text{\ \ for\ \ } l_{n+1}=a, N_{n+1}=e$
                                    }; \\
                                    
                                    & \node(d2){$\q\prl\qip{n} \text{~}\equiv\text{~} \q\prl a[e]\prl a[e]\prl \qip{n}$};&\node{\tc1};\\
                                    & \node    {${}= \q\prl \pim{n+2}{N}\text{\ \ for\ \ }
                                    l_{n+1}=l_{n+2}=a,\ 
                                    N_{n+1}=N_{n+2}=e$
                                    };\\
                                    
                                    & \node(d3){$\q\prl\qip{n}$};&\node{\tc1}; \\
                                    
  \node (rootq) {$\q\prl\qip{n}$};  & \node(d5){$\q\prl\qip{n}$};&\node{\tc1}; \\
                                    
                                    & \node(d7){$\q\prl\qip{n} \text{~}\equiv\text{~} \q\prl a[e]\prl\qip{n} $};&\node{\tc1}; \\
                                    & \node    {${}= \q\prl\pim{n+1}{N} \text{\ \ for\ \ }
                                    l_{n+1}=a,\ N_{n+1}=e$
                                    }; \\
                                    
                                    & \node(d11){$\q\prl a[0]\prl\qip{n-1} \text{~}=\text{~} \q\prl\pimp{n}{N'}$};&\node{\tc1}; \\
                                    & \node    {\text{~for~}
                                    $l_i'=l_i$,
                                    $\ N_i'=N_i~(0\le i\le n-1)$,
                                    $\ l_n'=a$,
                                    $\ N_n'=0$
                                    }; \\
                                    
                                    & \node(d12){$\q\prl a[0]\prl\qip{n-1} \text{~}=\text{~} \q\prl\pimp{n}{N'}$};&\node{\tc1}; \\
                                    & \node     {\text{~for~}
                                    $l_i'=l_i$,
                                    $\ N_i'=N_i~(0\le i\le n-1)$,
                                    $\ l_n'=a$,
                                    $\ N_n'=0$
                                    }; \\
  };

  \draw  [->, bend left ] (rootp) to  (c1.west) ;
  \draw  [->, bend left ] (rootp) to  (c2.west) ;
  \draw  [->, bend left ] (rootp) to  (c3.west) ;
  \draw  [->,           ] (rootp) to  (c5.west) ;
  \draw  [->, bend right] (rootp) to  (c7.west) ;
  \draw  [->, bend right] (rootp) to  (c11.west) ;
  \draw  [->, bend right] (rootp) to  (c12.west) ;

  \draw [->, bend left, dashed ] (rootq) to (d1.west) ;
  \draw [->, bend left, dashed ] (rootq) to (d2.west) ;
  \draw [->, bend left, dashed ] (rootq) to (d3.west) ;
  \draw [->,          , dashed ] (rootq) to (d5.west) ;
  \draw [->, bend right,dashed ] (rootq) to (d7.west) ;
  \draw [->, bend right]         (rootq) to (d11.west) ;
  \draw [->, bend right]         (rootq) to (d12.west) ;

}
  \caption{Simulation of internal transitions (dotted lines mean zero transitions)}
  \label{fig:internal}
\end{figure}

In a similar way, we explain how $\tau$ transitions of $P$ are matched by
$Q$, with another pair of transitions trees described in Figure~\ref{fig:internal}.

(1) When an $a[e \prl \ba e]$ from $P_0$ turns into $a[0]$, $Q$ does not have to do any
action, for we work with weak bisimulations. By replication, $Q$ can produce a copy $a[e]$
(or alternatively $a[\ba e]$) from $Q_0$, and since $(0,e)$ is in $\E$, we can add the
$a[0]$ and the copy $a[e]$ to the products $\prod$;
(2) $P$ can also make a reaction between two copies of $a[e \prl \ba e]$ in $P_0$, leaving
behind $a[e]$ and $a[\ba e]$. As in (1), $Q$ can draw two copies of $a[e]$ from $Q_0$, and
each product can be enlarged by two elements;
(3) it is also possible for $M_n=e\prl{}\ba e$ to do a $\tau$ transition, becoming
$M_n'=0$. It stands that $(M_n',N_n)\in\E$ and we are done;
(4) very similarly, two processes $M_n$ and $M_{n-1}$ may react, becoming $M_n'$ and
$M_{n-1}'$. It stands also that $(M_{n-1}',N_{n-1})$ and $(M_n',N_n)$ are in $\E$, so the
resulting processes are still related;
(5) it is possible for $M_n$ to follow the transition $M_n\lts{\alpha}M_n'$ and react with
a copy from $P_0$ which leaves behind $a[\alpha]$ (since $\ba \alpha$ has been consumed to
conclude the reaction). Again, it stands that $M_n'$ and $N_n$ are related by $\E$, and
that we can draw an $a[e]$ from $Q_0$ to pair it with the residue $M_n'$ in the products
$\prod$;
(6) also, a copy $a[e \prl \ba e]$ from $P_0$ may passivate an $l_i[M_i]$, provided $l_i =
e$, and leave a residue $a[\ba e]$. $Q$ can do the same passivation using $Q_0$'s $a[e]$,
and leave $a[0]$.  As it happens that $(\ba e,0)$ is in $\E$, the residues can be added to
the products too;
(7) finally, the process $l_n[M_n]$, if $l_n = e$, may be passivated by $M_{n-1}$,
reducing the size of $P$'s product. $Q$ can passivate $l_n[N_n]$ too, using a copy $a[e]$
from $P_0$, which becomes $a[0]$ after the reaction. $Q$'s product too is shorter, but we
need to add the $a[0]$ to it. To do so, we draw a copy $a[e \prl \ba e]$ from $P_0$, and
since $(e \prl \ba e,0)$ is in $\E$, $a[e \prl \ba e]$ and $a[0]$ are merged into their
respective product.

This ends the sketch of the proof that $\X$ is an environmental bisimulation, and
therefore that $\bang{}a[e \prl \ba e]$ and $\bang{}a[e] \prl a[\ba e]$ are behaviourally
equivalent.

\subsection{Overview of the paper}
\label{sub:Overview of the paper}

The rest of this paper is structured as follows. In Section~\ref{sec:HOpiP} we describe
the higher-order $\pi$-calculus with passivation. In Section~\ref{sec:EB} we formalize our
environmental bisimulations.  In Section \ref{sec:Examples} we give some examples of
bisimilar processes.  In Section~\ref{sec:conc}, we bring up some future work to conclude
our paper.

% LATEXMAKE gappei.tex
% LATEXEXTRA -- vim extra
% vim: set textwidth=90 spelllang=en_gb spell:
\section{Higher-order $\pi$-calculus with passivation} %
\label{sec:HOpiP}
We introduce a slight variation of the higher-order $\pi$-calculus with
passivation~\cite{hopip}---HO$\pi$P for short---through its syntax and a labelled
transitions system.

\subsection{Syntax} 
\label{sub:Syntax}
The syntax of our HO$\pi$P processes $P$, $Q$ is given by the following grammar, very
similar to that of Lenglet \emph{et al.}~\cite{hopip} (the higher-order $\pi$-calculus
extended with located processes and their passivation):
\[
\begin{array}{lcl}
  P, Q &::=& 0 ~\mid~ a(X).P ~\mid~ \ba a\langle M\rangle.P ~\mid~ (P \prl P) ~\mid~ a[P]
  ~\mid~ \nu a.P ~\mid~ {\bang{}P} ~\mid~ \run(M)\\
  M, N &::=& X ~\mid~ `P
\end{array}
\]
$X$ ranges over the set of variables, and $a$ over the set of names which can be used for
both locations and channels. $a[P]$ denotes the process $P$ running in location $a$.  To
define a general up-to context technique (Definition~\ref{def:ebuptc}, see also
Section~\ref{sec:conc}), we distinguish terms $M$, $N$ from processes $P$, $Q$ and adopt
explicit syntax for processes as terms $`P$ and their execution $\run(M)$.

\subsection{Labelled transitions system}
\label{sub:Labelled transitions system}
We define $\mathit{n}$, $\mathit{fn}$, $\mathit{bn}$ and $\mathit{fv}$ to be the functions
that return respectively the set of names, free names, bound names and free variables of a
process or an action. We abbreviate a (possibly empty) sequence $x_0,x_1,\dots,x_n$ as
$\ti x$ for any meta-variable $x$.  The transition semantics of HO$\pi$P is given by the
following labelled transition system, which is based on that of the higher-order
$\pi$-calculus (omitting symmetric rules \textsc{Par-R} and \textsc{React-R}):
\begin{small}
\begin{center}
$\inferrule*[Right=Ho-in]{ }{a(X).P \lts{a(M)} P\{M/X\}}$\qquad \qquad \qquad
$\inferrule*[Right=Ho-out]{ }{\ba a\langle M\rangle.P \lts {\ba a\langle
M\rangle} P}$\\
~\\
$\inferrule*[Right=Par-l]{P_1 \lts{\alpha}P'_1 \quad \mathit{bn}(\alpha)\cap \mathit{fn}(P_2) = \emptyset}{P_1 \prl P_2 \lts{\alpha} P'_1  \prl  P_2}$\qquad\qquad\qquad
$\inferrule*[Right=Rep]{!P \prl P \lts{\alpha} P'} {!P \lts{\alpha}P'}$\\
~\\
$\inferrule*[Right=React-l]{P_1 \lts{(\nu \ti b).\ba a\langle M\rangle} P'_1 \quad P_2 \lts{a(M)} P'_2 \quad \{\ti b\} \cap \mathit{fn}(P_2) = \emptyset}{P_1 \prl P_2 \lts {\tau} \nu \ti b. (P'_1 \prl P'_2)}$\\
~\\
$\inferrule*[Right=Guard]{P \lts{\alpha} P' \quad a \not \in \mathit{n}(\alpha)} {\nu a.P \lts{\alpha} \nu a.P'}$ \qquad \qquad \qquad
$\inferrule*[Right=Extr] {P\lts{(\nu \ti b). \ba a\langle M\rangle}P'\quad c\neq a\quad c\in \mathit{fn}(M)\setminus \{\ti b\}}
{\nu c. P \lts{\nu (\ti b,c).\ba a\langle M\rangle}P'}$\\
\end{center}
\end{small}
extended with the following three rules:
{\small
\[
\inferrule*[Right=Transp]{P \lts{\alpha} P'}{a[P] \lts{\alpha} a[P']} \qquad \qquad \qquad
  \inferrule*[Right=Passiv]{ }{a[P] \lts{\ba a\langle`P\rangle} 0} \qquad \qquad \qquad
\inferrule*[Right=Run]{ }{\run (`P) \lts{\tau} P}
\]
}

Assuming again knowledge of the standard higher-order $\pi$-calculus~\cite{hop,pisan}, we
only explain below the three added rules that are not part of it.  The \texttt{Transp}
rule expresses the \emph{transparency} of locations, the fact that transitions can happen
below a location and be observed outside its boundary. The \texttt{Passiv} rule
illustrates that, at any time, a process running under a location can be passivated
(stopped and turned into a term) and sent along the channel corresponding to the
location's name.  Quotation of the process output reminds us that higher-order
communications transport terms.  Finally, the \texttt{Run} rule shows how, at the cost of
an internal transition, a process term be instantiated.  As usual with small-steps
semantics, transition does not progress for undefined cases (such as $\run(X)$) or when
the assumptions are not satisfied.

Henceforth, we shall write $\ba a.P$ to mean $\ba a\langle`0\rangle.P$ and $a.P$ for
$a(X).P$ if $X\not\in\mathit{fv(P)}$. We shall also write $\eqv$ for the structural
congruence, whose definition is standard (see the appendix,
%Definition~\ref{def:A:struct-cong}).
Definition~A.1).

% LATEXMAKE gappei.pdf
% LATEXEXTRA -- vim extra
% vim: set textwidth=90 spelllang=en_gb spell:
\section{Environmental bisimulations of HO$\pi$P}
\label{sec:EB}
Given the higher-order nature of the language, and in order to get round the universal
quantification issue of context bisimulations, we would like observations (terms) to be
stored and reusable for further testing.  To this end, let us define an
\emph{environmental relation} $\X$ as a set of elements $(r,\E,P,Q)$ where
  $r$ is a finite set of names,
  $\E$ is a binary relation (with finitely many free names) on variable-closed terms
  (i.e.\  terms with no free variables), and
  $P$ and $Q$ are variable-closed processes.

We generally write $\adj{x}{S}$ to express the set union $\{x\}\cup S$.  We also
use graphically convenient notation $\EB{{P}}{{Q}}{{r}}{{\E}}$ to mean
$(r,\E,P,Q) \in \X$ and define the \emph{term context closure} $\Clos =
\E \cup \{(`P,`Q) \mid (P,Q)\in \pClos\}$ with the \emph{process context closure} $\pClos
=\{(C[\M],C[\N])\mid\M\E\N,\ C\textit{ context},\ \mathit{bn}(C)\cap
\mathit{fn}(\E,r)=\emptyset,\ \mathit{fn}(C)\subseteq r\}$, where a context is a
process with zero or more \emph{holes} for terms.
Note the distinction of terms $`P$, $`Q$ from processes $P$, $Q$.
We point out that $\zClos$ is the identity on terms with free names in
$r$, that $\Clos$ includes $\E$ by definition, and that the context closure
operations are monotonic on $\E$ (and $r$).  Therefore, for any $\E$ and $r$, the
set $\Clos$ includes the identity $\zClos$ too.
Also, we use the notations $\fst{\mathcal S}$ and $\snd{\mathcal S}$ to denote
the first and second projections of a relation (i.e.~set of pairs) $S$.  Finally, we define weak
transitions $\Lts{}$ as the reflexive, transitive closure of $\lts{\tau}$, and
$\Lts{\alpha}$ as $\Lts{}\lts{\alpha}\Lts{}$ for $\alpha \not= \tau$ (and define
$\Lts\tau$ as $\Lts{}$).

We can now define environmental bisimulations formally:

{\definition
\label{def:eb}
An environmental relation $\X$ is an environmental bisimulation if $\EB{P}{Q}{r}{\E}$
implies:
\begin{enumerate}
  \item\label{1} if $P \lts{\tau} P'$, then $\exists Q'.~Q \Lts{} Q'$ and $\EB{P'}{Q'}{r}{\E}$,

  \item\label{2} if $P \lts{a(M)} P'$ with $a\in r$, and if $(M,N)\in\Clos$, then
    $\exists Q'.~ Q \Lts{a(N)} Q'$ and $\EB{P'}{Q'}{r}{\E}$,

  \item\label{3} if $P \lts{\nu \ti b.\ba a\langle M\rangle} P'$ with $a\in r$ and $\ti
    b\not\in \mathit{fn}(r,\fst{\E})$,
    then $\exists Q', N.~Q \Lts{\nu \ti c.\ba a\langle N\rangle} Q'$
    with $\ti c\not\in \mathit{fn}(r,\snd{\E})$ and $\EB{P'}{Q'}{r}{\adj{(M,N)}\E}$,

  \item\label{4} for any $(`P_1,`Q_1) \in \E$ and $a\in r$, we have $\EB{P \prl a[P_1]}{Q \prl a[Q_1]}{r}{\E}$,

  \item\label{5} for any $n \not \in \mathit{fn}(\E,P,Q)$, we have
    $\EB{P}{Q}{\adj n r}{\E}$, and

  \item\label{6} the converse of 1, 2 and 3 on $Q$'s transitions.
\end{enumerate}
}
Modulo the symmetry resulting from clause~\ref{6}, clause~\ref{1} is usual; clause~\ref{2}
enforces bisimilarity to be preserved by any input that can be built from the knowledge,
hence the use of the context closure; clause~\ref{3} enlarges the knowledge of the
observer with the leaked out terms.  Clause~\ref{4} allows the observer to spawn (and
immediately run) terms concurrently to the tested processes, while clause~\ref{5} shows
that he can also create fresh names at will.

A few points related to the handling of free names are worth mentioning: as the set
of free names in $\E$ is finite, clause~\ref{5} can always be
applied; therefore, the attacker can add arbitrary fresh names to the set $r$ of
known names so as to use them in terms $M$ and $N$ in clause~\ref{2}.
Fresh $\ti b$ and $\ti c$ in clause~\ref{3} also exist thanks to the finiteness of
free names in $\E$ and $r$.

We define environmental bisimilarity $\sim$ as the union of all environmental
bisimulations, and it holds that it is itself an environmental bisimulation (all the
conditions above are monotone on $\X$).  Therefore, $\Eq{P}{Q}{r}{\E}$ if and only if
$\EB{P}{Q}{r}{\E}$ for some environmental bisimulation $\X$. We do particularly
care about the situation where $\E=\emptyset$ and $r=\mathit{fn}(P,Q)$. It corresponds to
the equivalence of two processes when the observer knows all of their free names (and thus
can do all observations), but has not yet learnt any output pair.

For improving the practicality of our bisimulation proof method, let us devise an
up-to context technique~\cite[p. 86]{pisan}: for an environmental relation $\X$, we write $\cEB{P}{Q}{r}{\E}$
if $P \eqv \nu\ti c.(P_0 \prl P_1)$, $Q \eqv \nu\ti d.(Q_0 \prl Q_1)$, $\EB{P_0}{Q_0}{r'}{\E'}$,
$(P_1,Q_1) \in \gpClos{r'}{\E'} $,
$\E \subseteq \gClos{r'}{\E'}$, $r\subseteq r'$, and $\{\ti
c\}\cap \mathit{fn}(r,\fst{\E})=\{\ti d\}\cap \mathit{fn}(r,\snd{\E}) =
\emptyset$.
As a matter of fact, this is actually an up-to context and up-to environment and up-to
restriction and up-to structural congruence technique, but because of the clumsiness of
this appellation we will restrain ourselves to ``up-to context'' to preserve clarity.  To
roughly explain the convenience behind this notation and its (long) name:
(1)~``up-to context'' states that we can take any $(P_1,Q_1)$ from the
(process) context closure
$\gpClos{r'}{\E'}$ of the environment $\E'$ (with free names in $r'$) and execute them in
parallel with processes $P_0$ and $Q_0$ related by $\X_{\E';r'}$;
similarly, we allow environments $\E$ with terms that are not in $\E'$ itself but are in the (term) context closure $\gClos{r'}{\E'}$;
(2)~``up-to environment'' states that, when proving the bisimulation clauses,
we please ourselves with environments $\E'$ that are \emph{larger} than the $\E$ requested by Definition~\ref{def:eb};
(3)~``up-to restriction'' states that we also content ourselves with tested processes $P$, $Q$ with
extra restrictions $\nu \ti c$ and $\nu \ti d$ (i.e.~less observable names);
(4)~finally, ``up-to structural congruence'' states that we identify all processes that
are structurally congruent to $\nu \ti c.(P_0 \prl P_1)$ and $\nu \ti d.(Q_0 \prl Q_1)$.

Using this notation, we define environmental bisimulations up-to context as
follows:

{\definition
\label{def:ebuptc}
An environmental relation $\X$ is an environmental bisimulation up-to context if
$\EB{P}{Q}{r}{\E}$ implies:
\begin{enumerate}
  \item\label{11} if $P \lts{\tau} P'$, then $\exists Q'.~Q \Lts{} Q'$ and
    $\cEB{P'}{Q'}{r}{\E}$,

  \item\label{22} if $P \lts{a(M)} P'$ with $a\in r$, and if $(M,N)\in\Clos$,
    then $\exists Q'.~Q \Lts{a(N)} Q'$ and $\cEB{P'}{Q'}{r}{\E}$,

  \item\label{33} if $P \lts{\nu \ti b.\ba a\langle M\rangle} P'$ with $a\in r$ and $\ti
    b\not\in \mathit{fn}(r,\fst{\E})$,
    then $\exists Q',N.~Q \Lts{\nu \ti c.\ba a\langle N\rangle} Q'$
    with $\ti c\not\in \mathit{fn}(r,\snd{\E})$ and $\cEB{P'}{Q'}{r}{\adj{(M,N)} \E}$,

  \item\label{44} for any $(`P_1,`Q_1) \in \E$ and $a\in r$, we have
    $\cEB{P \prl a[P_1]}{Q \prl a[Q_1]}{r}{\E}$,

  \item\label{55} for any $n \not \in \mathit{fn}(\E,P,Q)$, we have
    $\EB{P}{Q}{\adj n r}{\E}$, and

  \item\label{66} the converse of 1, 2 and 3 on $Q$'s transitions.
\end{enumerate}
}
The conditions on each clause (except~\ref{55}, which is unchanged for
the sake of technical convenience) are weaker than that
of the standard environmental bisimulations, as we require in the positive instances
bisimilarity modulo a context, not just bisimilarity itself. It is important to remark
that, unlike in~\cite{EB} but as in~\cite{hoapp}, we do not need a specific context to
avoid stating a tautology in clause~\ref{44}; indeed, we spawn terms $(`P_1,`Q_1)\in\E$
immediately as processes $P_1$ and $Q_1$, while the context closure can only use the terms
under an explicit $\run{}$ operator.

We prove the soundness (under some condition; see Remark~\ref{rem:simple}) of
environmental bisimulations as follows.  Full proofs are found in the appendix,
%Section~\ref{A:EBProofs} but are nonetheless sketched below.
Section~B but are nonetheless sketched below.

\begin{lemma}[Input lemma]\label{in}
  If $(P_1,Q_1)\in\pClos$ and $P_1\lts{a(M)}P_1'$ then $\forall N.\exists Q_1'.$
$Q_1\lts{a(N)}Q_1'$ and $(P_1',Q_1')\in{\epClos{\adj{(M,N)}{\E}}}$.
\end{lemma}

\begin{lemma}[Output lemma]\label{out}
If $(P_1,Q_1)\in\pClos$, $\{\ti b\} \cap \mathit{fn}(\E,r)=\emptyset$ and $P_1\lts{\nu\ti
b.\ba a\langle M\rangle}P_1'$ then $\exists Q_1',N.$ $Q_1\lts{\nu\ti b.\ba a\langle
N\rangle}Q_1'$, $(P_1',Q_1')\in\rpClos{\adj{\ti b}r} $ and $(M,N)\in\rClos{\adj{\ti b}r}$.
\end{lemma}

\begin{definition}[Run-erasure]
  \label{def:run-erasure}
  We write $P \le Q$ if $P$ can be obtained by (possibly repeatedly)
  replacing zero or more subprocesses $\run(`R)$ of $Q$ with $R$, and
  write $\mEBY{P}{Q}{r}{\E}$ for $\cEBY{P\le}{\ge Q}{r}{\le\E\ge}$.
\end{definition}

\begin{definition}[Simple environment]
  A process is called \emph{simple} if none of its subprocesses has the
  form $\nu a.P$ or $a(X).P$ with $X \in \mathit{fv}(P)$.  An
  environment is called simple if all the processes in it are simple.
  An environmental relation is called simple if all of its
  environments are simple (note that the tested processes may still be
  non-simple).
\end{definition}

\begin{lemma}[Reaction lemma]\label{tau}
  For any simple environmental bisimulation up-to context $\Y$, if
  $\mEBY{P}{Q}{r}{\E}$ and $P \lts\tau P'$, then there is a $Q'$ such
  that $Q \Lts\tau Q'$ and $\mEBY{P'}{Q'}{r}{\E}$.
\end{lemma}
{\proofsketch{Lemmas~\ref{A:in},~\ref{A:out} and~\ref{A:tau}}
Lemma \ref{in} (\resp \ref{out}) is proven by
straightforward induction on the transition derivation of
$P_1\lts{a(M)}P_1'$ (\resp $P_1\lts{\nu\ti b.\ba a\langle M\rangle}P_1'$).
Lemma~\ref{tau} is proven last, as it uses the other two lemmas (for the internal
communication case).
}

\begin{lemma}[Soundness of up-to context]
  \label{lem:sound-uptc}
Simple bisimilarity up-to context is included in bisimilarity.
\end{lemma}
{\proofsketch{%IGNORED
}
By checking that $\{(r,\E,P,Q)\mid\mEBY{P}{Q}{r}{\E}\}$ is included in $\sim$,
where $\Y$ is the simple environmental bisimilarity up-to context. In particular, we use Lemma~\ref{in} for
clause~\ref{2}, Lemma~\ref{out} for clause~\ref{3}, and Lemma~\ref{tau} for clause~\ref{1}
of the environmental bisimulation.
}

Our definitions of reduction-closed barbed equivalence $\approx$ and
congruence $\simcong$ are standard and omitted for brevity; see
%the appendix, Definition~\ref{def:A:redbarb} and~\ref{def:A:redcong}.
the appendix, Definition~B.2 and~B.3

\begin{theorem}[Barbed equivalence from environmental bisimulation]\label{thm:redbarb}
  \\
If $\mEBY{P}{Q}{\mathit{fn}(P,Q)}{\emptyset}$ for a simple
  environmental bisimulation up-to context $\Y$, then $P \approx Q$.
\end{theorem}
{\proofsketch{}
By verifying that
each clause of the definition of $\approx$ is implied by membership of $\Y^-$, using
Lemma~\ref{lem:sound-uptc} for the parallel composition clause.}

\begin{corollary}[Barbed congruence from environmental bisimulation]
  \label{cor:congruence}
  \\
  If $\mEBY{\ba a\langle`P\rangle}{\ba
    a\langle`Q\rangle}{\adj{a}{\mathit{fn}(P,Q)}}{\emptyset}$ for a
  simple environmental bisimulation up-to context $\Y$, then $P
  \simcong Q$.
\end{corollary}
We recall that, in context bisimulations, showing the equivalence of
$\ba a \langle `P \rangle$
and $\ba a \langle `Q \rangle $ almost amounts to testing the equivalence of $P$ and $Q$ in
every context. However, with environmental bisimulations, only the location context in
clause~\ref{4} of the bisimulation has to be considered.

\begin{remark}\label{rem:simple}
  The extra condition ``simple'' is needed because of a technical
  difficulty in the proof of Lemma~\ref{tau}: when an input process
  $a(X).P$ is spawned under location $b$ in parallel with an output
  context $\nu c.\ba a\langle M\rangle.Q$ (with $c \in
  \mathit{fn}(M)$), they can make the transition $b[a(X).P
  \prl \nu c.\ba a\langle M\rangle.Q] \lts{\tau} b[\nu c.(P\{M/X\}
  \prl Q)]$, where the restriction operator $\nu c$ appears
  \emph{inside} the location $b$ (and therefore can be passivated
  together with the processes); however, our spawning clause only
  gives us $b[a(X).P] \prl \nu c.\ba a\langle M\rangle.Q \lts{\tau}
  \nu c.(b[P\{M/X\}] \prl Q)$ and does not cover the above case.
  Further investigation is required to overcome this difficulty (although we have not yet found a concrete
  counterexample of soundness, we conjecture some modification to the
  bisimulation clauses would be necessary).  Note that, even if the
  environments are simple, the tested processes do not always have to
  be simple, as in Example~\ref{ex:6} and \ref{ex:7}.  Moreover,
  thanks to up-to context, even the output terms (including passivated
  processes) can be non-simple.
\end{remark}

\section{Examples} %

% LATEXMAKE gappei.tex
% LATEXEXTRA -- vim extra
% vim: set textwidth=90 spelllang=en_gb spell:
\label{sec:Examples}
Here, we give some examples of HO$\pi$P processes whose behavioural equivalence is proven
with the help of our environmental bisimulations.  In each example, we prove the
equivalence by exhibiting a relation $\X$ containing the two processes we consider, and by
showing that it is indeed a bisimulation up-to context (and environment, restriction and
structural congruence).
We write $P \prl \dots \prl P$ for a finite, possibly null, product of the process $P$.

% LATEXMAKE gappei.tex
\begin{example}
\label{ex:1}
$e \prl \bang{}a[e] \prl \bang{}a[0] \approx\bang{}a[e] \prl \bang{}a[0]$.
(This example comes from~\cite{hopip}.)
\end{example}
\proof Take
$\X =  \{(r,~\emptyset,~e \prl P,~P) \mid r \supseteq \{a,e\}\}
\cup
\{(r,\emptyset,P,P)\mid r \supseteq \{a,e\}\}$
where
$P = \bang{}a[e] \prl \bang{}a[0]$.
It is immediate to verify that whenever $P\lts{\alpha}P'$, we have $P'\equiv P$, and
therefore that transition $e \prl P\lts{\alpha}e \prl P'\equiv e \prl P$ can be matched by
$P\lts{\alpha}P'\equiv P$ and conversely.  Also, for $e \prl P\lts{e}P$, we have that
$P\lts{e}\bang{}a[e]\prl{}a[0] \prl\bang{}a[0]\equiv P$ and we are done since
$(r,\emptyset,P,P)\in \X$.  Moreover, the set $r$
must contain the free names of $P$, and to satisfy clause~\ref{55} about adding fresh
names, bigger $r$'s must be allowed too.  The passivations of $a[e]$ and $a[0]$ can be
matched by syntactically equal actions with the pairs of output terms $(`e,`e)$ and
$(`0,`0)$ included in the identity, which in turn is included in the context closure
$\zClos$.  Finally clause~\ref{44} of the bisimulation is vacuously satisfied because the
environment is empty.
We therefore have $e \prl \bang{}a[e] \prl \bang{}a[0] \approx \bang{}a[e] \prl
\bang{}a[0]$ from the soundness of environmental bisimulation up-to context.

% LATEXMAKE gappei.tex
\begin{example}
\label{ex:2}
$\bang{}\ba a \prl \bang{}e~\approx \bang{}a[e]$.
\end{example}
%{\proofsketch{Example~\ref{ex:A:2}} Take
{\proofsketch{} Take
$\X =  \{(r,~\E,~P,~Q)\mid r \supseteq \{a,e,l_1,\dots,l_n\}\mid$
$\E=  \{(`0,`e)\}$,
$\ n\ge 0$,
$\ P =  \bang{}\ba a \prl  \bang{}e \prl \prod_{i=1}^{n}l_i[0]$, 
$\ Q =  \bang{}a[e] \prl      \prod_{i=1}^{n}l_i[e] \prl a[0] \prl  \dots \prl a[0]\}$.
See the appendix, Example~C.1 for the rest of the proof.
}

% LATEXMAKE gappei.tex
\begin{example}
\label{ex:5}
$\bang{}a[e] \prl \bang{}b[\ba e] \approx \bang{}a[b[e \prl \ba e]]$.
This example shows the equivalence proof of more complicated processes
with nested locations.
\end{example}
{\proofsketch{} Take:
\[
\begin{array}{lclllll}
  \X & = & \{(r,~\E,~P,~Q)\mid & r &\multicolumn{2}{l}{\supseteq\{a,e,b,l_1,\dots,l_n\},}\\
  &&&P_0&= \bang{}a[e] \prl \bang{}b[\ba e],&~Q_0= \bang{}a[b[e \prl \ba e]],\\
  &&&P  &\multicolumn{2}{l}{=P_0\prl \prod_{i=1}^n l_i[P_i] \prl b[0]\prl \dots\prl b[0],}\\
  &&&Q  &\multicolumn{2}{l}{=Q_0\prl \prod_{i=1}^n l_i[Q_i],}\\
  &&&\multicolumn{3}{l}{(`\ti P,`\ti Q)\in \E,~n\ge 0\},}\\
  \E &=& \multicolumn{5}{l}{
  \{(`x,`y)\mid x\in\{0,e,\ba e\},~ y\equiv\in\{0,e,\ba e,(e \prl \ba e),b[0],b[e],b[\ba e],b[e \prl \ba e]\}\}}.
\end{array}
\]
See the appendix, Example~C.2 for the rest of the proof.
}

% LATEXMAKE gappei.tex
\begin{example}
\label{ex:6}
$c(X).\run(X) \approx \nu f.(f[c(X).\run(X)] \prl \bang{}f(Y).f[\run(Y)])$.
The latter process models a system where a process $c(X).\run(X)$ runs in location $f$, and
executes any process $P$ it has received. In parallel is a process $f(Y).f[\run(Y)]$ which
can passivate $f[P]$ and respawn the process $P$ under the same location $f$. Informally,
this models a system which can restart a computer and resume its computation after a
failure.
\end{example}
{\proof
Take $\X    =  \X_1 \cup \X_2$ where:
\[
\begin{array}{lcllllllll}
  \X_1 & = & \multicolumn{8}{l}{\{(r,~\emptyset,~ c(X).\run(X),~\nu f.(f[c(X).\run(X)]\prl\bang{}f(Y).f[\run(Y)]))\mid r\supseteq\{c\}\},}\\
  \X_2 & = & \{(r,~\emptyset,~P,~Q)\mid
  &r&\supseteq&\adj{c}{\mathit{fn}(R)},&\hskip0.5em S& =& \run(`\run(\dots`\run(`R)\dots)),\\
  &&&P&\in&\{\run(`R),R\}, &\hskip0.5em Q& = & \nu \ti
  f.(f[S]\prl\bang{}f(Y).[\run(Y)])
  \}.
\end{array}
\]
As usual, we require that $r$ contains at least the free name $c$ of the tested processes.
All outputs belong to $\zClos$ since they come from a process $R$ drawn from $\zClos$, and
therefore, we content ourselves with an empty environment $\emptyset$.  Also, by the
emptiness of the environment, clause~\ref{44} of environmental bisimulations is vacuously
satisfied.

Verification of transitions of elements of $\X_1$, i.e.\ inputs of some $`R$ (with
$(`R,`R)\in\zClos)$ from $c$, is immediate and leads to checking elements of $\X_2$.  For
elements of $\X_2$, we observe that $P=\run(`R)$ can do one $\tau$ transition to become
$R$, while $Q$ can do an internal transition passivating the process $\run(`R)$ running in
$f$ and place it inside $f[\run(`~)]$, again and again. $Q$ can also do $\tau$
transitions that consume all the $\run(`~)$'s until it becomes $R$. Whenever $P$ (\resp
$Q$) makes an observable transition, $Q$ (\resp  $P$) can consume the $\run(`~)$'s and
weakly do the same action as they exhibit the same process. We observe that all
transitions preserve membership in $\X_2$ (thus in $\X$), and therefore we have that $\X$
is an environmental bisimulation up-to context, which proves the behavioural equivalence
of the original processes $c(X).\run(X)$ and $c(X).\nu
f.(f[c(X).\run(X)]\prl\bang{}f(Y).f[\run(Y)])$.
}

% LATEXMAKE gappei.tex
\begin{example}
\label{ex:7}
$c(X).\run(X) \approx c(X).\nu a.(\ba a\langle X\rangle\prl \bang{}\nu
f.(f[a(X).\run(X)] \prl f(Y).\ba a \langle Y \rangle ))$.
This example is a variation of Example~\ref{ex:6} modelling a system where computation is
resumed on another computer after a failure.
\end{example}
{\proof
Take $\X = \X_1 \cup \X_2\cup \X_3$ where:
\[
\begin{array}{lcll}
  \X_1 & = & \multicolumn{2}{l}{\{(r,~\emptyset,~c(X).\run(X),~c(X).\nu a.(\ba a\langle X\rangle\prl F))\mid r\supseteq\{c\}\},}\\

  \X_2 & = & \multicolumn{2}{l}{\{(r,~\emptyset ,~P_1,~\nu a.(F \prl R_1 \prl
  R_2 \prl \ba a\langle `P_2\rangle))\mid}\\
  &&&r\supseteq\adj{\{c\}}{\mathit{fn}(P)},\hskip0.5em P_1,P_2 \in
  \{\run(`P),P\},\hskip0.5em R_1=\ba a\langle N_1\rangle \prl \dots \prl \ba a\langle
  N_n\rangle,\\
  &&&R_2 = \nu l_1.(l_1[Q_1]\prl l_1(Y).\ba a\langle Y\rangle) \prl \dots \prl \nu
  l_m.(l_m[Q_m]\prl l_m(Y).\ba a\langle Y\rangle),\\
  &&&N_1,\dots,N_n,`Q_1,\dots,`Q_m = `\run(`\run(\dots`\run(`a(X).\run(X))\dots)),\ n\ge 0\},\\

  \X_3 & = & \multicolumn{2}{l}{\{(r,~\emptyset ,~P_1,~\nu a.(F \prl R_1 \prl
  R_2 \prl \nu l.(l[P_2] \prl l(Y).\ba a\langle Y\rangle)))\mid}\\
  &&&r\supseteq\adj{\{c\}}{\mathit{fn}(P)},\hskip0.5em P_1,P_2 \in
  \{\run(`P),P\},\hskip0.5em R_1=\ba a\langle N_1\rangle \prl \dots \prl \ba a\langle
  N_n\rangle,\\
  &&&R_2 = \nu l_1.(l_1[Q_1]\prl l_1(Y).\ba a\langle Y\rangle) \prl \dots \prl \nu
  l_m.(l_m[Q_m]\prl l_m(Y).\ba a\langle Y\rangle),\\
  &&&N_1,\dots,N_n,`Q_1,\dots,`Q_m = `\run(`\run(\dots`\run(`a(X).\run(X))\dots)),\ n\ge 0\},\\

  F    & = & \multicolumn{2}{l}{\bang{}\nu f.(f[a(X).\run(X)]\prl f(Y).\ba a\langle Y\rangle).}\\
\end{array}
\]
\newcommand{\lhs}{\emph{lhs}}
\newcommand{\rhs}{\emph{rhs}}
The set of names $r$ and the environment share the same fate as those of
Example~\ref{ex:6} for identical reasons.  For ease, we write \lhs{} and \rhs{} to
conveniently denote each of the tested processes.

Verification of the bisimulation clauses of $\X_1$ is immediate and leads to a member
$(r,\emptyset,\run(`P),\linebreak[0]\nu a.(\ba a \langle`P \rangle \prl F))$ of $\X_2$ for some $`P$
with $(`P,`P)\in\zClos$.  For $\X_2$, \lhs{} can do an internal action (consuming its
outer $\run(`~)$) that \rhs{} does not have to follow since we work with weak
bisimulations, and the results is still in $\X_2$; conversely, internal actions of \rhs{}
do not have to be matched.  Some of those transitions that \rhs{} can do are reactions
between replications from $F$. All those transitions creates elements of either $R_1$ or
$R_2$ that can do nothing but internal actions and can be ignored further in the proof
thanks to the weakness of our bisimulations.

Whenever \lhs{} does an observable action $\alpha$, that is, when $P_1 = P\lts{\alpha}P'$,
\rhs{} must do a reaction between $\ba a \langle `P_2 \rangle $ and $F$, giving $\nu
l.(l[P_2] \prl l(Y).\ba a \langle Y \rangle )\Lts{\alpha } \nu l.(l[P'] \prl l(Y).\ba a
\langle Y \rangle)$ which satisfies $\X_3$'s definition. Moreover, all transitions of
$P_1$ or $P_2$ in $\X_3$ can be matched by the other, hence preserving the membership in
$\X_3$.  Finally, a subprocess $\nu l.(l[P_2] \prl l(Y).\ba a \langle Y \rangle) $ of
\rhs{} of $\X_3$ can do a $\tau$ transition to $\ba a \langle `P_2 \rangle$ and the
residues belong back to $\X_2$.

This concludes the proof of behavioural equivalence of the original processes
$c(X).\run(X)$ and $c(X).\nu a.(\ba a\langle X\rangle.\bang{}\nu f.(f[a(X).\run(X)] \prl
f(Y).f[\run(Y)]))$. 
}

% LATEXMAKE gappei.tex
% LATEXEXTRA -- vim extra
% vim: set textwidth=90 spelllang=en_gb spell:
\section{Discussion and future work}
\label{sec:conc}
In the original higher-order $\pi$-calculus with passivation described by Lenglet \emph{et
al.} ~\cite{hopip}, terms are identified with processes: its syntax is just $P ::= 0 \mid
X \mid a(X).P \mid \ba a\langle P\rangle.P \mid (P \prl{} P) \mid a[P]   \mid   \nu a.P
\mid   {\bang{}P}$.  We conjecture that it is also possible to develop sound environmental
bisimulations (and up-to context, etc.) for this version of HO$\pi$P, as we~\cite{EB} did for
the standard higher-order $\pi$-calculus.
However we chose not to cover directly the original higher-order $\pi$-calculus with
passivation, for two reasons: (1) the proof method of~\cite{EB} which relies on guarded
processes and a factorisation trick using the spawning clause of the bisimulation is
inadequate in the presence of locations; (2) there is a very strong constraint in
clause 4 of up-to context in~\cite[Definition E.1 (Appendix)]{EB} (the context has no hole
for terms from $\E$).  By distinguishing processes from terms, not only is our up-to
context method much more general, but our proofs are also direct and technically simple.
Although one might argue that the presence of the $\run$ operator is a burden, by using
Definition~\ref{def:run-erasure},
one could devise an ``up-to $\run$'' technique and abstract
$\run(\dots`\run(`P))$ as $P$, making equivalence proofs easier to write and understand.

\newcommand{\pqcont}{\bang b[\outpz{a}]}
\newcommand{\outp}[2]{\ba {#1} \langle #2 \rangle }
\newcommand{\outpz}[1]{\ba{#1}}
As described in Remark~\ref{rem:simple}, removing the limitation on the environments is left for future work.
We also plan to apply environmental bisimulations to (a substantial subset of) the Kell
calculus so that we can provide a practical alternative to context bisimulations in a more
expressive higher-order distributed process calculus. In the Kell calculus, locations are
not transparent: one discriminates messages on the grounds of their origins (i.e.\  from a
location above, below, or from the same level). For example, consider the (simplified)
Kell processes $P = \outp{a}{M}.\pqcont$ and $Q = \outp{a}{N}.\pqcont$ where $M=\ba a$ and
$N=0$. They seem bisimilar assuming environmental bisimulations naively like those in
this paper: intuitively, both $P$ and $Q$ can output (respectively $M$ and $N$) to channel
$a$, and their continuations are identical; passivation of spawned $l[M]$ and
$l[N]$ for known location $l$ would be immediately matched; finally, the output to
channel $a$ under $l$, turning $P$'s spawned $l[M]$ into $l[0]$, could be matched by an
output to $a$ under $b$ by $Q$'s replicated $b[\outpz{a}]$.
However, $M$ and $N$ behave differently when observed from the same level (or below),
for example as in $l[M \prl a(Y).\ba{ok}]$ and $l[N \prl a(Y).\ba{ok}]$ even under the presence
of $\pqcont$.
More concretely, the context $[\cdot]_1 \prl a(X).c[X \prl a(Y) . \ba{ok}]$
distinguishes $P$ and $Q$, showing the unsoundness of such naive definition.  This
suggests that, to define sound environmental bisimulations in Kell-like calculi with
non-transparent locations, we should require a stronger condition such as
bisimilarity of $M$ and $N$ in the output clause. Developments on this idea are in
progress.

\bibliography{gappei}
\bibliographystyle{abbrv}
%\newpage
%\input{appendix}

\newpage

\end{document}